\documentstyle[preprint,aps,floats,epsf,tighten,eqsecnum]{revtex}
\begin{document}
\draft
\preprint{
 \parbox{1.5in}{\leftline{JLAB-THY-98-13}
                \leftline{WM-98-101} }  }

\title{ Pole Term and Gauge Invariance in Deep Inelastic Scattering }

\author{Zolt\'an Batiz}
\address{ Department of Physics, College of William and Mary,
 Williamsburg, VA  23187}
\author{Franz Gross}
\address{Department of Physics, College of William and Mary, Williamsburg, 
VA 23187 \\Thomas Jefferson National Accelerator Facility, Newport
News, VA 23606}
\date{\today}
\maketitle

\begin{abstract}

In this paper we reconcile two contradictory statements about deep
inelastic scattering (DIS) in manifestly covariant theories: (i) the
scattering must be gauge invariant, even in the deep inelastic limit, and
(ii)  the pole term (which is not gauge invariant in a covariant theory)
dominates the scattering amplitude in the deep inelastic limit. An
``intermediate''  answer is found to be true.  We show that, at all
energies, the gauge dependent part of the pole term cancels the
gauge dependent part of the rescattering term, so that both the pole
and rescattering terms can be separately redefined in a gauge invariant
fashion. The resulting, redefined pole term is then shown to dominate the
scattering in the deep inelastic limit.  Details are worked out for a simple
example in 1+1 dimensions.

\end{abstract}

\widetext

\section{Introduction}
Deep inelastic scattering (DIS) is a very important tool for investigating 
the structure of hadrons\cite{West}. It involves the interaction between an
energetic lepton beam (electrons, muons, neutrinos) and the respective
hadron.  

DIS has been described (i) in the framework of 
the Quark Parton Model (QPM) using the light-cone formalism and (ii) with the
aid of covariant field theories. The QPM is based on the assumption that all
quarks (before and after scattering) are on-shell \cite{Roberts}. In this
manner gauge invariance is built-in. The leading contribution is the pole
(or Born) term and the rescattering term is negligable in the deep
inelastic limit. In descriptions based on covariant field theory, however,
the intermediate quarks (or nucleons in studies of the
deuteron\cite{covardeut}) are off-shell, and hence the pole term cannot, by
itself, be gauge invariant.  In this case it would appear that the pole term
cannot be the leading term in the deep inelastic limit, and this has been
a problem for covariant theories for many years.  

To avoid this problem, de Forest \cite{df83} replaced the current by
its gauge invariant part.  He presented no proof, and the ``de Forest
prescription''  has always seemed adhoc. The problem continues to trouble
calculations which rely on the use of off-shell particles and dominance of
the pole term.  In a recent paper Kelly \cite{Kelly} outlines three
alternatives.  Assuming that the photon has four momentum
$q^\mu=(\nu,0,0,q)$, these three alternatives can be summarized as follows:

\begin{itemize}

\item {\it de Forest prescription\/}: leave the time component of the current
(the charge) unchanged, and replace the third component by $J^3\to(\nu/q)
J^0$.

\item {\it Weyl prescription\/}: leave the third component of the current 
unchanged, and replace the time component by $J^0\to (q/\nu) J^3$.

\item {\it Landau prescription\/}: replace the four current by $J^\mu\to J^\mu
+ (J\cdot q/Q^2) \;q^\mu$,

\end{itemize}   

\noindent where we use the SLAC convention that $Q^2=-q^2>0$ for electron
scattering. Note that each of these descriptions gives {\it a different
result for the charge and $J^3$ components\/}.  In the generalized Breit
frame, where $q^\mu=(0,0,0,Q)$, the de Forest and the Landau prescriptions
are identical.  One of the major conclusions of this paper is a detailed
theoretical argument justifying the Landau prescription.  

In addition to this, we present a simple toy model for the structure
functions of a composite system.  The structure functions we obtain from
this model exhibit scaling, are dominated in the deep inelastic limit by the
``gauge invariant part'' of the pole term, and give a simple, qualitative
description of the distribution function of valence quarks.  We present
the results from two versions.  In one, the particles are taken to be
scalars, and the scalar ``nucleon'' is a bound state of a scalar ``quark''
with unit charge and a scalar diquark with no charge.  In a second, more
realistic model, the composite spin 1/2 nucleon is a bound state of a spin
1/2 quark and a scalar diquark.  Stimulated by the success of QCD in 1+1
dimensions\cite{Einhorn}, we carry out these model calculations in 1+1
dimensions, obtaining finite results without the need for model dependent
form factors.  The only parameters in the model are the masses of the quark
and diquark, $m_1$ and $m_2$ respectively.  The 1+1 dimensional calculations
do not necessarily give the right physics, but help in developing the
necessary tools for a more complete treatment.  

The remainder of this paper is divided into five sections.  First we present
our simple bound state model for the nuclon (which could also be applied to
the description of mesons).  Then we study gauge invariance and show how to
uniquely define the gauge invariant part of the pole term.  In Sec.~IV we
show for scalar quarks that the gauge invariant part of the pole term does
indeed dominate DIS.  The model deep inelastic structure functions are
calculated and compared with experiment in Sec.~V, and conclusions are given
in Sec.~VI.

\section{Bound State Equation for the ``Nucleon''}

We begin the discussion by constructing a dynamical model for the
nucleon.  In the first subsection we  will  
assume that the scalar ``nucleon'' is a bound state of the two
``fundamental''  scalar particles: a charged ``quark'' $q$ of mass
$m_1$ and an uncharged ``diquark'' $d$ of mass $m_2$. In the second subsection
we will generalize to a charged spin 1/2 quark and a charged spin 1/2 nucleon.

\begin{figure}
\centerline{\epsfxsize=2in\epsffile{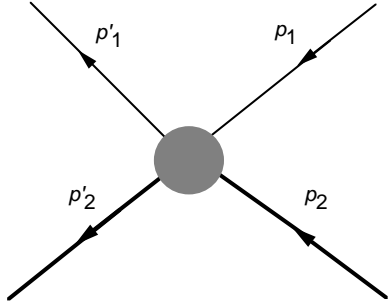}}
\caption{Elementary $qd$ interaction.}\label{one}
\end{figure}

\subsection{Scalar ''nucleons'' as bound states of scalar constituents}

The two "fundamental" constituents will interact with each other via the
coupling shown in Fig.~1:
\begin{equation}
V=\lambda f(p^2_2)f(p^{ \prime 2}_2)
\label{1eq1}
\end{equation}
where $\lambda$ is the coupling strength, and the form factor
$f(p_2^2)$ is a function of the square of the diquark four-momentum. 
For generality, we keep these form factors for now, but they will be set
to unity later when we carry out calculations in 1+1 dimensions.
Hence the scattering amplitude is:
\begin{equation}
{\cal M}= \lambda f(p^2_2)f(p^{\prime 2}_2) + (-i)^4 \lambda^2 
f(p^2_2)f(p^{\prime 2}_2)\,b(s) + ... =
\frac{\lambda f(p^2_2)f(p^{\prime 2}_2)}{1- \lambda\, b(s)};
\label{1eq2}
\end{equation}
%

\begin{figure}
\centerline{\epsfxsize=3in\epsffile{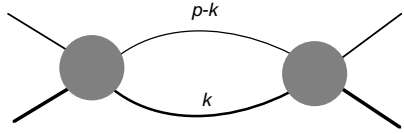}}
\vspace*{-0.4in}
\caption{The bubble diagram.}\label{two}
\end{figure}

\noindent where we have introduced the function $b(s)$ defined by
\begin{eqnarray}
b(s)=i \int {\frac{d^dk}{(2 \pi)^d} \frac{f^2(k^2)}{(m_2^2-k^2-i 
\epsilon)(m_1^2-(p-k)^2 -i \epsilon)}}\, ,
\label{1eq4}
\end{eqnarray}
with $p=p_1+p_2$ and $s=p^2$.  This ``bubble'' integral is represented
graphically in Fig.~\ref{two}.  The mass of the bound state must be below
threshold, and for simplicity we assume $m_2-m_1<M<m_1+m_2$. 

\begin{figure}[b]
\vspace*{-0.4in}
\centerline{\epsfxsize=3in\epsffile{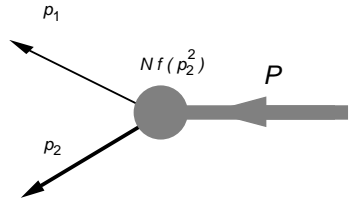}}
\caption{The ``nucleon''$qd$ vertex.}\label{four}
\end{figure}

\begin{figure}
\centerline{\epsfxsize=2.5in\epsffile{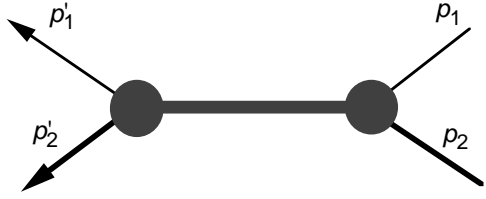}}
\caption{The dressed bound state pole contribution to the $qd$
scattering amplitude.}\label{three}
\end{figure}

In this simple separable model, the bound state is described by the
``nucleon''$qd$ vertex function
\begin{equation}
\Gamma(p_2)={\cal N} f(p_2^2)\, , \label{vertex}
\end{equation}
where ${\cal N}$ is a constant to be determined from the normalization of
the wave function.  This is shown in Fig.~\ref{four}. Denoting the dressed
propagator of the bound nucleon by $\Delta_B(p)$, the dressed $qd$
scattering amplitude is given exactly in this simple model by
\begin{equation}
{\cal M}=-\Gamma(p_2)\Delta_B(p) \Gamma(p'_2) \, , \label{scatamp}
\end{equation}
where
\begin{equation}
-{\cal N}^2 \Delta_B(p)= \frac{ \lambda}{1- \lambda \, b(s)}\, .
\label{1eq6}
\end{equation}
Eq.~(\ref{scatamp}) is illustrated in Fig.~\ref{three}.

Since the propagator should have a pole at $s=M^2$, it is useful to 
expand the bubble $b(s)$ in this vicinity, so the denominator of the
RHS of Eq.~(\ref{1eq6}) is 
$$1-\lambda b(M^2)+ \lambda (M^2-s)\,b'(M^2)+ \cdots$$
and this must vanish at $s=M^2$.   Consequently $\lambda b(M^2)=1$ is a
necessary condition of the bound state, and near the pole Eq.~(\ref{1eq6})
becomes
\begin{equation}
-{\cal N}^2 \Delta_B(p)= \frac{ 1}{(M^2-s)\,b'(M^2)+ \cdots}.
\label{1eq61}
\end{equation}
Knowing that, in the vicinity of $s=M^2$, the nucleon propagator should 
be
\begin{equation}
\Delta_B(p) \simeq \frac{1}{M^2-s- i \epsilon} \, ,
\end{equation}
allows us to find the normalization of the  $Nqd$ vertex function:
\begin{equation}
{\cal N}= \sqrt{\frac{-1}{b'(M^2)}} \, .
\label{1eq8}
\end{equation}
Note that the existance of the bound state implies that
$b'(M^2)<0$, which will be shown below. 

In 1+1 D the bubble integral will converge without form factors, and we
can assume, for simplicity, that $f(p^2_2) \equiv 1$.  The  bubble
integral (\ref{1eq4}) is then model independent, and easily calculated. 
If it is parametrized using  Feynman variables, 
one can either compute the loop below treshold directly in Minkowski space
(doing the momentum integration first and then integrating over the
Feynman variable) or compute it in Euclidian space and Wick rotate back
to the Minkowski space.  Introducing the triangle function 
\begin{equation}
\Delta(p^2, m_1^2, m_2^2)=(p^2-m_1^2-m_2^2)^2-4m_1^2m_2^2 \, ,
\label{triangle}
\end{equation}
which is negative if $(m_2-m_1)^2<p^2<(m_2+m_1)^2$, the bubble becomes:
\begin{equation}
b(s) = - \frac{1}{2 \pi} \frac{1}{\sqrt {- \Delta (s, m_1^2, m_2^2)}}\;
\arctan
\left[\frac{ \sqrt{- \Delta( s, m_1^2, m_2^2)}}{s-m_1^2-m_2^2}\right]\, .
\label{1eq9}
\end{equation}
This is real, and if the mass of the bound state is to be $M$,
then $\lambda$ must be
\begin{equation}
\lambda= -2 \pi\sqrt{- \Delta(M^2, m_1^2, m_2^2)}
 \left[ \arctan \left( \frac{ \sqrt{- \Delta(M^2, m_1^2,
m_2^2)}}{M^2- m_1^2 -m_2^2} \right) \right]^{-1}\,  .
\label{1eq10}
\end{equation}
This is real (so the Lagrangian is hermitian, as it should be) 
and negative (as expected from the fact that we have a bound state,
implying the interaction is attractive).  In fact, the whole 1+1 D theory
can be derived from the Lagrangian
\begin{equation}
{\cal L}= \frac{1}{2}( \partial^{\mu} d) ( \partial_{\mu} d) - 
\frac{1}{2} m_2^2 d^2 +  (\partial^{\mu} q ) (\partial_{\mu} q )^{\dagger} - m_1^2
qq^{\dagger} - 
\frac{\lambda}{2} d^2 qq^{\dagger},
\label{1eq13}
\end{equation}
where $q$ is the ``quark'' and $d$ the ``diquark'' fields.
From this Lagrangian, or from Eq. (\ref{1eq1}), one sees that $\lambda$
must be negative in order for a bound state to exist.

Above threshold, however, the loop $b(s)$ also has an imaginary part
\begin{equation}
\Im b(s) = \frac{-1}{2 \sqrt{ \Delta (s, m_1^2, m_2^2)}}\, ,
\label{1eq11}
\end{equation}
which can be calculated in two ways. The first method is a computation 
based on Cutkosky's rules, which are easily handled in this simple case.
The second follows directly from the
Feynman parametrization of Eq.~(\ref{1eq4}).  Integrating over the
internal momentum gives
\begin{equation}
b(s)=- \frac{1}{4 \pi} \int^{1}_{0}  \frac{dx}{m_1^2x + m_2^2(1-x) - 
s\, x(1-x)- i\epsilon} \, . 
\label{1eq14}
\end{equation}
Introducing the variable $s'= {m_2^2}/{x} +{m_1^2}/({1-x})$  transforms
(\ref{1eq14}) into:
\begin{equation}
b(s)= - \frac{1}{ \pi} \int^{\infty}_{(m_1+m_2)^2}  \frac{ds'}{s'-s -i
\epsilon}\; \left[\frac{1}{2 \sqrt{\Delta (s', m_1^2, m_2^2)}}\right] \, .
\label{1eq15}
\end{equation}
This expression yields a real value for $b(s)$ if $s$ is below 
threshold because $\Delta (s', m_1^2, m_2^2)$ is positive above threshold.
This relation is a dispersion relation, and comparing it with the
general form
\begin{equation}
f(s)= \frac{1}{\pi} \int^{\infty}_{\rm treshold} 
{ \frac{ds'\,\Im f(s')}{s'-s-i \epsilon}}\, ,
\label{1eq16}
\end{equation}
one concludes that the imaginary part of the bubble is correctly given
in Eq.~(\ref{1eq11}). As a byproduct, we have also
discovered the correct phase space factor for use with dispersion
relations in 1+1 D. 

The real part of $b(s)$ above treshold can also be found in
two ways: (i) using Eq. (\ref{1eq16}), or (ii) by analytically
continuing Eq. (\ref{1eq9}), a procedure which also provides a third
method for computing the bubble's imaginary part. As a result of these
calculations, the real part above treshold is found to be:
\begin{equation}
\Re b(s)= \frac{-1}{4 \pi \sqrt{ \Delta (s, m_1^2, m_2^2)}}
\ln \left[ \frac{s^2 - s(m_1^2 + m_2^2) + s \sqrt{ \Delta (s, m_1^2,
m_2^2)}} {s^2 - s (m_1^2 + m_2^2) - s \sqrt{ \Delta (s, m_1^2,
m_2^2)}} \right]\, .
\label{1eq17}
\end{equation}
At large $s$, $\Im b(s)  \sim s^{-1}$, $\Re b(s)  \sim  s^{-1} 
\ln({s}/{m_2^2})$; so $b(s) \rightarrow 0$ as $s \rightarrow
\infty$.  This important result will be used below.

Before turning to the spin 1/2 case, we use the representation
(\ref{1eq14}) to write the normalization condition (\ref{1eq8}) in the
following form:
\begin{equation}
1= \frac{{\cal N}^2}{4 \pi} \int^{1}_{0} dx \frac{x(1-x)}{\left[m_1^2x +
m_2^2(1-x) -  M^2\, x(1-x)\right]^2} \, . 
\label{1eqlast}
\end{equation}
This relation will be used again in the following sections. 

\subsection{Generalization to spin 1/2 nucleons and quarks}

In this subsection we will repeat our former calculations for the more
realistic case of charged spin 1/2 nucleons and quarks (retaining,
however, the uncharged scalar diquark).  The quark-diquark contact interaction
is assumed to be a scalar in the Dirac space of the quarks, with the momentum
space dependence previously given in Eq. ~(\ref{1eq1}).

In this case the bubble in $d$ dimensions becomes:
\begin{equation}
b(p)=i \int \frac{d^d k}{(2 \pi)^d} \frac{f^2(k^2)}{(m_2^2-k^2-i \epsilon)(
m_1-(\not p - \not k) -i \epsilon)},
\label{1eq18}
\end{equation}
or, after the integration:
\begin{equation}
b(p)= A(p^2) + B(p^2) \not p.
\label{1eq19}
\end{equation}
In 1+1 dimensions with the form factors set to unity, the factors $A(p^2)$
and $B(p^2)$ become:
\begin{eqnarray}
A(p^2)=&&-\frac{m_1}{4 \pi} \int^1_0 \frac{dx}{m_1^2x+m_2^2(1-x)-p^2x(1-x)}
\nonumber\ \\
B(p^2)=&&-\frac{1}{4 \pi} \int_0^1 \frac{(1-x)
\,dx}{m_1^2x+m_2^2(1-x)-p^2x(1-x)}\, ,
\label{1eq20}
\end{eqnarray}
where $x$ is the Feynman parameter used previously in Eq.~(\ref{1eq14}).
In the presence of spin, Eq.~(\ref{1eq2}) is modified:
\begin{equation}
{\cal M}=\frac{\lambda\; (1-\lambda A(p^2)+\lambda B(p^2) \not p)}{(1-\lambda
A(p^2))^2-\lambda^2B^2(p^2)p^2)}\, .
\label{1eq21}
\end{equation}
Expanding the denominator near the nucleon pole gives
\begin{eqnarray}
(1-\lambda A(p^2))^2-\lambda^2 B^2(p^2) p^2= && 
(1-\lambda
A_0)^2-2\lambda A'(1-\lambda A_0 )(p^2 - M^2) \nonumber   \\
&&-\lambda ^2 (B_0^2+2 B_0 B' (p^2-M^2))p^2+ \cdots\, ,
\label{1eq22}
\end{eqnarray}
where $A_0=A(p^2)$, $B_0=B(p^2)$, the primes denote derivatives with
respect of $p^2$, and all functions are evaluated at $p^2=M^2$.
The denominator vanishes at $p^2=M^2$ if
\begin{equation}
1-\lambda A_0 = M \lambda B_0.
\label{1eq23}
\end{equation}
This is the bound state condition.
After imposing the bound state condition, in the vicinity of the poles the
scattering amplitude becomes:
\begin{equation}
{\cal M} \simeq \frac{M+ \not
p}{(M^2-p^2)(B_0+2MA'+2M^2B')}.
\label{1eq24}
\end{equation}
The normalization factor ${\cal N}$ is then:
\begin{equation}
{\cal N}^{-2}=-\left[2M(A'+MB')+B_0\right]\, .
\label{1eq25}
\end{equation}
Using the relations (\ref{1eq20}), and the identity
\begin{equation}
0=\int^1_0 dx\;
\frac{m_1^2 x^2 - m_2^2 (1-x)^2 }{\left[m_1^2x+m_2^2(1-x)-M^2x(1-x)\right]^2}.
\label{1eq27a}
\end{equation}  
we may rewrite the 
renormalization condition (\ref{1eq25}) in the following form:
\begin{equation}
1=\frac{{\cal N}^{2}}{4 \pi} \int^1_0 dx\;
\frac{x(M(1-x)+m_1)^2 }{\left[m_1^2x+m_2^2(1-x)-M^2x(1-x)\right]^2}.
\label{1eq27}
\end{equation}  
This relation will be needed later to verify baryon number conservation.

\section{Deep Inelastic Scattering and Gauge Invariance}

We now add electromagnetic interactions to the model, and study the
implications of the requirement that the e.m.~interaction be gauge
invariant.  We begin by restoring the form factors; later we will move
to 1+1 dimensions and set the form factors to unity.  As in the previous
section, we will first carry out the discussion for a scalar quark and
nucleon, and later extend it to a spin 1/2 quark and nucleon.

\subsection{The scalar model} 

\begin{figure}
\centerline{\epsfxsize=3in\epsffile{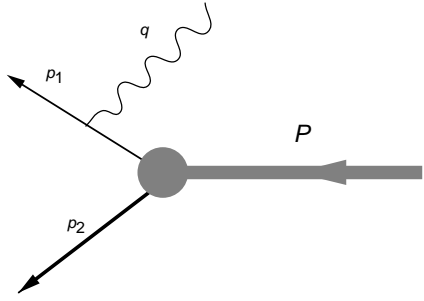}}
\caption{The pole term.}\label{five}
\end{figure}
  
Start with the pole term, shown in Fig.~\ref{five}, which gives
\begin{equation}
e{\cal J}^{ \mu}_a =i(-i)^3 e j^{ \mu} (p_1, p^{ \prime}_1)
\Delta(p_1^{\prime})
\Gamma (p_2)\, ,
\label{2eq1}
\end{equation}
where $p'_1=p_1-q$, the propagator of the scalar (charged) ``quark'' is
\begin{equation}
\Delta (p_1) = 
\frac{1}{m_1^2-p_1^2-i\epsilon}\, ,
\label{2eq200}
\end{equation}
and $j^{\mu}(p_1, p_1^{\prime})$ is the one-body
current of the quark.  (Note that our definition of the current ${\cal J}$
does {\it not\/} include the charge.)  We assume that the diquark has no
electromagnetic interaction, so there are no diagrams with the photon
coupling to the diquark.  We require that the quark current satisfy the
Ward-Takahashi (WT) identity:
\begin{equation}
q^{ \mu} j_{ \mu}(p_1, p_1^{ \prime}) = \Delta (p_1^{ \prime}) -  \Delta
(p_1)\, .
\label{2eq2}
\end{equation}
The current of a bare, spinless particle 
\begin{equation}
j^{\mu}(p_1, p_1^{ \prime}) = (p_1+  p'_1)^\mu 
\label{2eq2aa}
\end{equation}
satisfies this identity.

Using the WT identity, and noting that $\Delta^{-1}(p_1)=0$ because
$p_1^2=m_1^2$, we see that the pole term by itself is not gauge invariant
\begin{eqnarray}
q_{ \mu} {\cal J}^{ \mu}_a =&& - q_{ \mu} j^{ \mu} (p_1, p_1^{ \prime})
\Delta (p_1^{\prime}) \Gamma (p_2) \nonumber\\ 
=&& -\left[\Delta^{-1} (p_1^{ \prime}) - \Delta^{-1}
(p_1) \right] \,\Delta (p_1^{\prime}) \Gamma (p_2) \nonumber\\
=&& - \Gamma(p_2) \ne0\, . 
\label{2eq4}
\end{eqnarray}
Hence the pole term by itself cannot be a good
approximation for the scattering amplitude.

\begin{figure}
\centerline{\epsfxsize=4in\epsffile{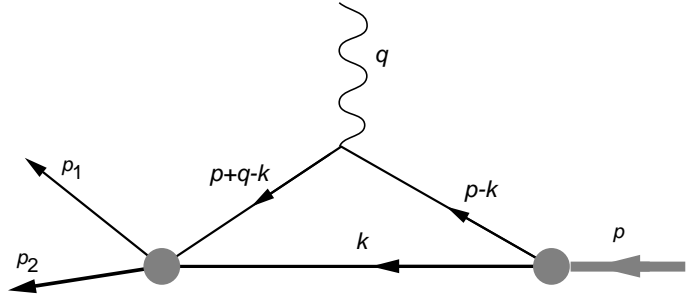}}
\caption{The rescattering diagram.}\label{six}
\end{figure}

To obtain a gauge invariant result, it is necessary to add the
rescattering term shown in Fig.~\ref{six}.  This term is
\begin{eqnarray}
e{\cal J}_b^{\mu}=&& i(-i)^6 e \frac{\lambda  f(p_2^2)}{1- \lambda b(s')}
\nonumber\\
&&\times\int \frac{d^dk}{(2
\pi)^d} 
\frac{f(k^2) j^{\mu} (p'-k, p-k)\Gamma(k)}{ (m_1^2 - (p'-k)^2 - i
\epsilon) (m_1^2 - (p-k)^2 - i \epsilon) (m_2^2 - k^2 -i \epsilon)}\, ,
\label{2eq5}
\end{eqnarray}
where $p'=p+q$ and $s^{ \prime} = p'^2$ is the square of the
momentum of the final state after it has absorbed the virtual
(space-like) photon.  Note that $p^2=s=M^2$ but that $p'^2=s'\ge M^2$.

This diagram is also not gauge invariant.  Using Eq.~(\ref{vertex}) and
the WT identity gives
\begin{equation}
q_{ \mu}{ \cal J}^{ \mu}_b = -\frac{{\cal N}f(p_2^2) }{1- \lambda  
b(s^{\prime})} \left[ \lambda b(s^{\prime}) - \lambda b(M^2)\right] =
{\cal N}f(p_2^2) = \Gamma(p_2)\, ,
\label{2eq6}
\end{equation}
where the bound state condition $\lambda b(M^2)=1$ was used in the second
step.  This is not gauge invariant, but if we add the
rescattering term to the pole term the total amplitude is
\begin{equation}
q_{ \mu} \left({\cal J}^{ \mu}_a + {\cal J}^{ \mu}_b\right) = 0 \, .
\end{equation}
Stated in a different way, the gauge dependent parts of the pole term
and the rescattering term cancel.

We now face the question posed in the introduction: How can the pole
term dominate the scattering in the deep inelastic limit if it is not
gauge invariant?  To answer this question we break the pole term into a
gauge invariant part and a gauge dependent part as follows:
\begin{eqnarray}
{\cal J}_a^\mu = -  {(2p_1-q)^{\mu} \Gamma (p_2) \over m_1^2-(p_1-q)^2}
= -  {\left[I_1^{\mu}(p_1) + I_2^\mu(p_1)\right]\,\Gamma (p_2) \over
m_1^2-(p_1-q)^2}
\, ,
\label{2eq8}
\end{eqnarray}
where
\begin{eqnarray}
I^{\mu}_1(p_1) =&& 2\left(p_1^{ \mu} - \frac{(q \cdot
p_1)}{q^2}q^{\mu}\right)
\, ,\label{2eq9} \\
I^{\mu}_2(p_1) =&& - q^\mu\left(1 - \frac{2(q \cdot p_1)}{q^2}\right)
=-{q^\mu\over q^2} \left[(p_1-q)^2-m_1^2\right] \, .
\label{2eq9a} 
\end{eqnarray}
Note that $q_\mu I_1^\mu=0$, and that the gauge dependent part of
the pole term reduces to
\begin{equation}
{\cal J}^{ \mu}_{ag} = -\frac{q^{ \mu}}{q^2}\,\Gamma (p_2) \, .
\label{2eq10}
\end{equation}

Now the rescattering term can also be similarly decomposed.  Noting that
\begin{equation}
j^{\mu} (p'-k, p-k) = I_1^\mu(p'-k) + I_2^\mu(p'-k) 
\end{equation}
the gauge dependent part of the rescattering term becomes
\begin{eqnarray}
{\cal J}^{\mu}_{bg}=&&-{q^\mu\over q^2 }\;\frac{\lambda \Gamma(p_2)}
{[1- \lambda b(s')]}   \nonumber\\
&&\times i\int \frac{d^dk}{(2 \pi)^d}
\frac{f^2(k^2)\Bigl\{(p-k+q)^2-(p-k)^2\Bigr\}}
{(m_2^2-k^2)(m_1^2-(p-k)^2)(m_1^2-(p-k+q)^2)}\nonumber\\
=&&{q^\mu\over q^2 }\,\Gamma(p_2)\;\frac{\lambda }
{[1- \lambda b(s')]} \Bigl\{ b(M^2)-b(s') \Bigr\} 
={q^\mu\over q^2 }\,\Gamma(p_2)
\label{2eq12}\, ,
\end{eqnarray}
which exactly cancels the gauge dependent part of the pole term!

We have shown that, if we drop the parts from each terms, the Born pole term
and the rescattering term are {\it redefined\/} so that each is {\it
separately gauge invariant\/}. As it turns out, the redefined pole term is
identical to the Landau prescription defined in the introduction.  We have
just justified this prescription.   

It is clear that the same decomposition works in 1 + 1 dimensions, where
the form factor $f=1$.  In this case the diagrams come
from from the  Lagrangian (\ref{1eq13}), modified  to include
electromagnetic interactions.  The new Lagrangian is
\begin{equation}
{\cal L} = \frac{1}{2} (\partial^{ \mu} d)(\partial_{ \mu} d) -\frac{1}{2} m_2^2 d^2 + (D_{
\mu} q)(D^{ \mu}q)^{\dagger} - m_1^2qq^{\dagger} - \frac{ \lambda}{2} 
d^2 qq^{\dagger} \, ,
\label{2eq13}
\end{equation}
where the fields are the same as in the preceding section, but the
gradient of the charged field has been replaced by the covariant
derivative
\begin{equation}
D^{ \mu} = \partial ^{ \mu} -ieA^{ \mu}\, .
\label{2eq14}
\end{equation}

\subsection{Extension to sin 1/2 quarks and nucleons}

We how extend the above results to the spin 1/2 model previously
introduced.  The Born term for a structureless quark (with the form factor
set equal to unity) is
\begin{equation}
e{\cal J}^{ \mu}_a =- e j^{ \mu} (p_1, p^{ \prime}_1)
S(p_1^{\prime})\Gamma (p_2,p)\, ,
\label{2beq1}
\end{equation}
where now $j^\mu=\gamma^\mu$ and the quark propagator is
\begin{equation}
S(p_1^{\prime}) = {1\over m_1 -\not p' -i\epsilon}  \, .
\label{2beq2}
\end{equation}
The vertex function for the spin 1/2 case is 
\begin{equation}
\Gamma (p,\lambda_N)={\cal N}{U}_N(p, \lambda_N) 
\end{equation}
where $\lambda_N$ is the helicity of the nucleon and ${U}_N(p,\lambda_N)$
are the helicity spinor of the nucleon. 
Using the Dirac equation for the final quark, the pole term can be written
\begin{equation}
{\cal J}^{\mu}=-\frac{{\cal N}}{m_1^2-(p_1-q)^2}\;{\bar u}(p_1, \lambda_1)
\left[ 2p_1^{\mu}- \gamma ^{\mu}\not q \right] {U}_N(p, \lambda_N),
\label{5eq1}
\end{equation}
where $\lambda _1$ the helicity of the outgoing quark and $u(p_1,\lambda_1)$
the helicity spinor of the quark. Now the current is spin dependent.

As before, we decompose the pole term into a gauge invariant and gauge
noninvariant part:
\begin{eqnarray}
{\cal J}^{\mu}_a=&&-\frac{{\cal N}}{m_1^2-(p_1-q)^2}\;
{\bar u}(p_1, \lambda_1)
\left[ 2p_1^{\mu}- \frac{2 p_1 \cdot q q^{\mu}}{q^2}- \gamma ^{\mu} \not q
+ q^{\mu} \right] {U}_N(p, \lambda_N) \nonumber\\
{\cal J}^{\mu}_{ag}=&&-\frac{ q^\mu}{q^2}\;{\cal N}
{\bar u}(p_1, \lambda_1) {U}_N(p, \lambda_N) = -\frac{
q^\mu}{q^2}\;{\bar u}(p_1, \lambda_1)\Gamma(p,\lambda_N)
\, .
\label{5eq1b}
\end{eqnarray}
Note that, except for the addition of the spinors, the gauge dependent part
identical to the scalar case, and the gauge independent part is again the
Landau prescription.

We now look at the spinor rescattering diagram.  Introducing
the decomposition 
\begin{eqnarray}
j^{\mu} (p'-k, p-k) =&& \gamma^\mu - {q^\mu \not q\over q^2} +{q^\mu \not
q\over q^2 }\nonumber\\
=&& \gamma^\mu - {q^\mu \not q\over q^2} - {q^\mu\over
q^2}\,\Bigl\{S^{-1}(p-k+q) - S^{-1}(p-k)\Bigr\} \, , 
\end{eqnarray}
the gauge dependent part of the rescattering term becomes
\begin{eqnarray}
{\cal J}^{\mu}_{bg}=&&{q^\mu\over q^2 }\;\lambda\;{\bar u}(p_1,
\lambda_1)  \left[1- \lambda b(p+q)\right]^{-1}   \nonumber\\
&&\times i\int \frac{d^dk}{(2 \pi)^d}
\frac{1}
{(m_2^2-k^2)} S(p-k+q)\Bigl\{S^{-1}(p-k+q) -S^{-1}(p-k)\Bigr\}\nonumber\\
&&\qquad\times S(p-k)
\;\Gamma(p,\lambda_N)   \nonumber\\ 
=&&{q^\mu\over q^2}\;{\bar u}(p_1,\lambda_1)  
\left[1- \lambda b(p+q)\right]^{-1}  \Bigl\{ \lambda\,b(p)-\lambda\,b(p+q)
\Bigr\}\;\Gamma(p,\lambda_N) \nonumber\\
=&&{q^\mu\over q^2 }\;{\bar u}(p_1,\lambda_1)\; \Gamma(p,\lambda_N)
\label{2xeq12}\, ,
\end{eqnarray}
where we used the bound state condition (\ref{1eq23}) and $\not p
\Gamma(p,\lambda_N) = M \Gamma(p,\lambda_N)$.  Note that this term exactly
cancels the gauge dependent part of the pole term!

This argument shows that this technique is quite general, and should
work in every case.  The pole term is divided into a term which is gauge
invariant and one proportional to $q^\mu$ by adding and subtracting a (unique)
term proportional to $q^\mu$.  Then it may be shown that this non gauge
invariant term is cancelled when the rescattering term is similarily
separated into (unique) gauge invariant and non guage invariant parts. 
Knowing that this works means that the non gauge invariant part of the pole
term {\it can be discarded without checking that it is cancelled by an
identical piece from the rescattering term\/}. Our argument can be
extended to currents with form factors if we use the technique developed by
Gross and Riska \cite{GR} to define the off-shell currents.

\section{Rescattering and The Triangle Diagram}

In this section we consider the rescattering term for the scalar model in more
detail. We  compute it exactly for the case $q=0$ and confirm that the charge
of the bound state is properly normalized.  Then we study the behavior of the
elastic form factor and the inelastic rescattering term at large $q^2$.  We
will compare the gauge invariant part of the rescattering term with the
gauge invariant part of the pole term and show that the pole term
dominates at high $q^2$.  In order to achieve convergence in a model
independent way, we return to 1+1 dimension.

The key to these calculations is the triangle diagram.  The gauge invariant
part of this diagram is 
\begin{equation}
t^{ \mu}(p',p)= -{2i\over q^2} \int \frac{d^2k}{(2
\pi)^2}\;\frac{q^2(p-k)^{\mu} - q^\mu\,q\cdot(p-k)}{(m_2^2 - k^2)(m_1^2 -
(p+q-k)^2)(m_1^2 - (p-k)^2)}  \, .
\label{3eq1}
\end{equation}
Feynman parametrizing the integral and doing the momentum integration gives
\begin{equation}
t^{ \mu}(p',p) = \left( p^{\mu} -{q^\mu\,q\cdot p\over
q^2}\right)\; \frac{1}{8 \pi}\int^{1}_{0} dx
\int^{1-x}_{0} dy \;\; \frac{1-x-y}{D^2(x,y,p,q)}\, ,
\label{3eq2a}
\end{equation}
where the denominator is
\begin{eqnarray}
D(x,y,p,q)=&&(px+qx+py)^2\nonumber\\
&& +x\,\left(m_1^2-(p+q)^2 -m_2^2\right) +
y\,\left(m_1^2-p^2 - m_2^2\right) + m_2^2 \, .
\end{eqnarray}
It is convenient to express the integral in terms of $p'=p+q$ and $p$, and to
change to the variables
\begin{eqnarray}
z=&&x+y \nonumber\\ 
u=&& {(x-y)\over z} \, ,
\label{3eq3}
\end{eqnarray}
which maps the region of integration into
\begin{eqnarray}
0&& \leq z \leq 1 \nonumber\\
- 1&& \leq u \leq 1 \, .
\label{3eq4}
\end{eqnarray}
The Jacobian of this transformation is $z/2$, and the integral then becomes
\begin{equation}
t^{ \mu}(p',p) = \left\{ (p'+p)^{\mu} -{q^\mu\,(p'^2-p^2)\over
q^2}\right\}\; \frac{1}{8 \pi}\int^{1}_{0} dz
\int^{1}_{-1} du \;\; \frac{z(1-z)}{d^2(z,u,p',p)}\, ,
\label{3eq2}
\end{equation}
where
\begin{eqnarray}
d(z,u,p',p) =&& m_2^2(1-z) + m_1^2z -{\textstyle{1\over2}}(p'^2+p^2) z(1-z)
\nonumber\\
&&-{\textstyle{1\over2}}(p'^2-p^2)uz(1-z)
-{\textstyle{1\over4}}q^2z^2(1-u^2)\, .
\label{d3eq2}
\end{eqnarray}
We will use the general results (\ref{3eq2}) and (\ref{d3eq2}) in the
following discussion.

\begin{figure}
\centerline{\epsfxsize=4in\epsffile{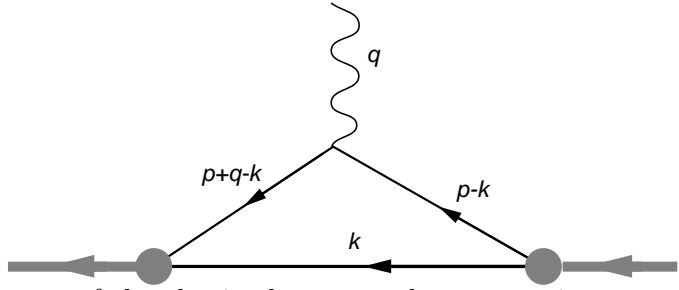}}
\caption{The hadronic part of the elastic electron-nucleon
scattering process, proportional to the deuteron form
factor.}\label{seven}
\end{figure}

\subsection{Charge Normalization}

First consider the hadronic part of the elastic electron-nucleon scattering
process, shown in Fig.~\ref{seven}.  This diagram is gauge invariant by
itself, and hence the deuteron form factor is given by
\begin{eqnarray}
e{\cal J}^\mu_e= e\,F(q^2)\,(p'+p)^\mu = e\,{\cal N}^2\,t^\mu(p',p) \, ,
\label{3eq8}
\end{eqnarray}
where, for elastic scattering, the triangle diagram is evaluated with
the constraints $p'^2 = (p+q)^2 = M^2=p^2$.  Substituting Eq.~(\ref{3eq2})
into Eq.~(\ref{3eq8}) gives
\begin{equation}
F(q^2) = \frac{{\cal N}^2}{8 \pi} \int^{1}_{0} dz
\int^{1}_{-1} du \;\; \frac{z(1-z)}{\left[\alpha (z)
-{\textstyle{1\over4}}q^2z^2(1-u^2)\right]^2}\, ,
\label{3eq5}
\end{equation}
where 
\begin{equation}
\alpha (z) = m_2^2 (1-z) + m_1^2z -M^2 z(1-z) \, .
\end{equation}  

If the charge is conserved, the form factor must satisfy the condition
\begin{eqnarray}
F(0)=1=&& \frac{{\cal N}^2}{8 \pi} \int^{1}_{0} dz
\int^{1}_{-1} du \;\; \frac{z(1-z)}{\left[m_2^2 (1-z) + m_1^2z -M^2
z(1-z)\right]^2}\nonumber\\
=&& {\cal N}^2\,{\partial\over \partial M^2}\left\{ \frac{1}{4 \pi}
\int^{1}_{0} dz\;\; \frac{1}{\left[m_2^2 (1-z) + m_1^2z -M^2
z(1-z)\right]} \right\} \label{Fat0}
\, ,
\end{eqnarray}
Comparing this with Eq.~(\ref{1eq14}) gives the relation
\begin{equation}
1=-{\cal N}^2 \frac{\partial b(M^2) }{\partial M^2} \, ,
\label{3eq7}
\end{equation}
which is in agreement with the Eq.~(\ref{1eq8}) and proves that charge is
conserved.

\subsection{Behavior of the Form Factor at Large Momentum Transfers}

In preparation for the discussion of rescattering in the deep inelastic
limit, and in order to study the properties of the 1+1 D model, we examine
the behavior of the form factor in the high $q^2$ limit.  This is obtained
from
\begin{eqnarray}
F(Q^2)=&& {\cal N}^2\,{\partial\over \partial M^2}\left\{ \frac{1}{8\pi}
\int^{1}_{0} dz\int^{1}_{-1} du\;\; \frac{1}{\left[\alpha(z)
 +{\textstyle{1\over4}}Q^2z^2(1-u^2)\right]} \right\}\nonumber\\
=&& {\cal N}^2\,{\partial\over \partial M^2}\left\{ \frac{1}{8\pi Q}
\int^{1}_{0} {dz\over z} \;\; {1\over\sqrt{\alpha(z) +
{\textstyle{1\over4}}Q^2z^2}}
\log\left({\sqrt{\alpha(z) + {\textstyle{1\over4}}Q^2z^2} + 
{\textstyle{1\over2}}Qz \over 
\sqrt{\alpha(z) + {\textstyle{1\over4}}Q^2z^2} - {\textstyle{1\over2}}Qz }
\right) \right\} \nonumber\\
=&& \frac{{\cal N}^2}{8\pi Q} \int^{1}_{0} (1-z) dz  \;\; {1\over
\alpha(z) \,\sqrt{\alpha(z) + {\textstyle{1\over4}}Q^2z^2}}\, ,
\end{eqnarray}
where we have introduced $Q^2=-q^2>0$ used in electron scattering.  At
large $Q$, the last integral is dominated by values of $z$ near zero, and is
well approximated by
\begin{eqnarray}
F(Q^2)\to &&  \frac{{\cal N}^2}{8\pi Q} \int^{1}_{0}  dz 
\;\; {1\over
\alpha(0) \,\sqrt{\alpha(0) + {\textstyle{1\over4}}Q^2z^2}}\nonumber\\
=&& \frac{{\cal N}^2}{4\pi Q^2}\int^{Q/2}_{0}  d\xi 
\;\; {1\over
m_2^2 \,\sqrt{m_2^2 + \xi^2}} \nonumber\\
=&&\frac{{\cal N}^2}{8\pi\, m_2^2\, Q^2} \;\;\log\left({\sqrt{4m_2^2 +
{Q^2}}  + {Q}\over
\sqrt{4m_2^2 +Q^2}  - {Q}} \right) \nonumber\\
\to&& \frac{{\cal N}^2}{4\pi\, m_2^2\, Q^2} \;\;\log\left( {Q\over m_2}
\right) \, .
\label{3eq16}
\end{eqnarray}
We see that $F(Q^2)$ falls off as $Q$ approaches infinity, as it should for a
composite particle.

\subsection{Implications for protons, neutrons, and mesons} 

These results can be applied to realistic cases.  Denoting the charge of the
quark by $e_q$ and the mass of the diquark by $m_2=m_d$ the total charge is
simply $\sum_i\,e_i$, as required by charge conservation.  At high
$Q^2$ however, we obtain
\begin{equation}
e\,F_N(Q^2)=\sum_i e_{q_i}\,F_{q_i}(Q^2)= \frac{{\cal N}^2}{4\pi\,
Q^2}\sum_i{e_{q_i}\over m_{d_i}^2}\log\left(\frac{Q}{m_{d_i}}\right)  \, .
\label{3eq188}
\end{equation}
Note that the sign of $F_p(Q^2)$ is  positive for the proton, but that it is
identically zero (or, more correctly, falls off faster than $Q^{-2}$) for the
neutron, {\it unless\/} the mass of the $ud$ diquark differs from the mass of
the $dd$ diquark.  In the latter case,
\begin{eqnarray}
e\,F_n(Q^2)=&&\left({2\over3}\right)\frac{e{\cal
N}^2}{4\pi\, Q^2}\left\{{1\over m_{dd}^2}\log\left(\frac{Q}{m_{dd}}\right) - 
{1\over m_{ud}^2}\log\left(\frac{Q}{m_{ud}}\right)\right\}\nonumber\\
e\,F_p(Q^2)=&&\left({2\over3}\right)\frac{e{\cal
N}^2}{4\pi\, Q^2}\left\{{2\over m_{ud}^2}\log\left(\frac{Q}{m_{ud}}\right) - 
{1\over2 m_{uu}^2}\log\left(\frac{Q}{m_{uu}}\right)\right\} \, .
\label{3eq189}
\end{eqnarray}
Ignoring the weak logarithimic dependence on the diquark mass, these are
equal (for example) if $m^2_{ud}=2m^2_{uu}=2m^2_{dd}$.  The behavior of these
elastic form factors at high $Q^2$ has not yet been determined
experimentally, and continues to be a subject of great interest. 

Note that Eq.~(\ref{3eq188}) also predicts that the electromagnetic form 
factors of spin 0 $q\bar q$  mesons which have the same flavor for
the quark and  antiquark (such as $s \bar s$, $c\bar c$), is identically zero.

\subsection{Rescattering in the deep inelastic limit}

\begin{figure}
\centerline{\epsfxsize=4in\epsffile{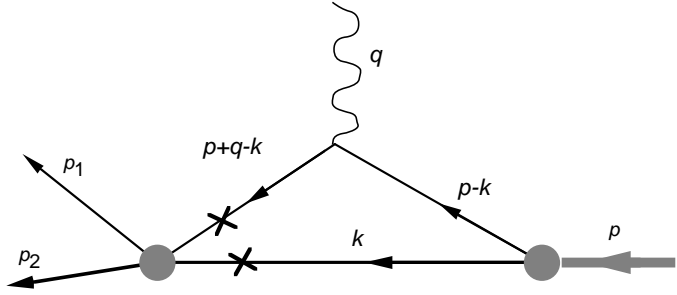}}
\vspace*{0.2in}
\caption{The two virtual particles in the final state can both be 
on shell (symbolized by the $\times$'s) if $W>m_1+m_2$.}\label{eight}
\end{figure}

In Sec.~III the rescattering term was treated briefly. We
divided the triangle into  a gauge dependent and gauge independent part.  The
former was calculated exactly, while the  latter is proportional to the
triangle diagram given in Eq.~(\ref{3eq2}):
\begin{eqnarray}
{\overline{\cal J}}_b^\mu =&& {\lambda{\cal N}\over \left[ 1-\lambda b(s')
\right] }\;t^\mu(p',p) \nonumber\\
=&& {\lambda{\cal N}\over \left[ 1-\lambda
b(W^2)\right]} \;\left\{ (p'+p)^{\mu} -{q^\mu\,(p'^2-p^2)\over
q^2}\right\}\; T(Q^2,W^2) \, ,
\end{eqnarray}
where $W^2=s'=p'^2$ and the scalar function $T(Q^2,W^2)$ is
\begin{equation}
T(Q^2,W^2) =  \frac{1}{8 \pi}\int^{1}_{0} dz
\int^{1}_{-1} du \;\;
\frac{z(1-z)}{\left[\alpha_W(z)+\beta(z,u)-i\epsilon\right]^2}
\, , \label{3eq2bb}
\end{equation}
where
\begin{eqnarray}
\alpha_W(z) =&&m_2^2(1-z)+m_1^2z-W^2z(1-z) \nonumber\\
\beta(z,u)=&&{\textstyle{1\over2}}(W^2-M^2) z(1-z)(1-u)
+{\textstyle{1\over4}}Q^2z^2(1-u^2)\, .
\label{d3eq2xx}
\end{eqnarray}
For electron scattering, $Q^2\ge0$ and $W^2-M^2\ge0$, so $\beta\ge0$. 
However, as for the bubble diagram, $\alpha_W$ has a zero if $W>m_1+m_2$.  
This means that the denominator of Eq.~(\ref{3eq2bb}) also has a zero, and
that
$T$ has an imaginary part. From the principles of dispersion theory we know
that this is due to the existance of physical scattering in the final state,
as illustrated in Fig.~\ref{eight}.

To obtain a tractible form for the integral, replace the $u$ integration by
$\beta$, which gives
\begin{equation}
T(Q^2,W^2) =  \frac{1}{4 \pi}\int^{1}_{0} dz
\int^{(W^2-M^2) z(1-z)}_{0} d\beta \;\;
\frac{z(1-z)}{\sqrt{\gamma(z,\beta)}
\,\left[\alpha_W(z)+\beta-i\epsilon\right]^2}
\, , \label{3eq2aa}
\end{equation}
where
\begin{eqnarray}
\gamma(z,\beta)=&&{\textstyle{1\over4}}[(W^2-M^2)z(1-z)+Q^2z^2]^2-Q^2z^2
\beta \, , \label{3eq2b}
\end{eqnarray}
We now replace $W^2$ by  
\begin{eqnarray}
W^2=(p+q)^2=M^2+Q^2\left({1\over x}-1\right)  \, 
\label{4eq3}
\end{eqnarray}
where $x$ is the Bjorken scaling variable
\begin{eqnarray}
x= {Q^2\over 2\,p\cdot q}= {Q^2\over 2\,M\nu} \, ,
\label{xdef}
\end{eqnarray}
and take the deep inelastic limit where $Q^2\to\infty$ with
$x$ held constant.  The integral is dominated by values of $z$ near zero,
and hence may be approximated by
\begin{eqnarray}
T(Q^2,W^2) \simeq &&  \frac{x}{2 \pi Q^2 (1-x)}\int^{1}_{0} dz
\int^{W^2 z}_{0} d\beta \;\;
\frac{1}{\left[m_2^2 -W^2z+\beta-i\epsilon\right]^2}\nonumber\\
\simeq  && \frac{x}{2 \pi Q^2 (1-x)}\int^{1}_{0} dz \left(
\frac{1}{\left[m_2^2 -W^2z -i\epsilon\right]} - \frac{1}{m_2^2} \right)
\nonumber\\
\simeq  && \frac{x}{2 \pi Q^2 (1-x)}\left( {i\pi x\over Q^2(1-x)}
 - \frac{2}{m_2^2} \right) \, , \label{tsim}
\end{eqnarray}
where we have kept the leading contribution from the imaginary part even
though it is negligible compared to the real part.

In conclusion, one can state that the gauge invariant part of the 
rescattering term falls  like $1/Q^2$.  Noting that the bubble also falls as
$1/Q^2$, the full rescattering term in the deep inelastic limit approaches 
\begin{eqnarray}
{\cal J}^{ \mu}_{b}\to-\left\{ (p'+p)^{\mu} -{q^\mu\,(p'^2-p^2)\over
q^2}\right\}\;\frac{\lambda{\cal N} x}{\pi\,m_2^2\, Q^2 (1-x)} \; 
\label{highres}
\end{eqnarray}
This is to be compared with the deep
inelastic limit of the Born term, which is
\begin{eqnarray}
{\cal J}^{ \mu}_{a}=&&-2\left(p_1^{ \mu}- \frac{(q
\cdot p_1)} {q^2}q^{ \mu}\right)\;\frac{{\cal N}}{m_1^2-(p_1-q)^2}\, ,
\label{3eq23}
\end{eqnarray}
and does not fall with $Q^2$.  Hence, in the deep inelastic limit the gauge
invariant part of the rescattering term is negligible in comparison to the
gauge invariant part of the pole term. 

In the next section we will calculate the cross section for deep inelastic
scattering, and find the structure functions predicted by this simple model.
%

\begin{figure}
\centerline{\epsfxsize=3in\epsffile{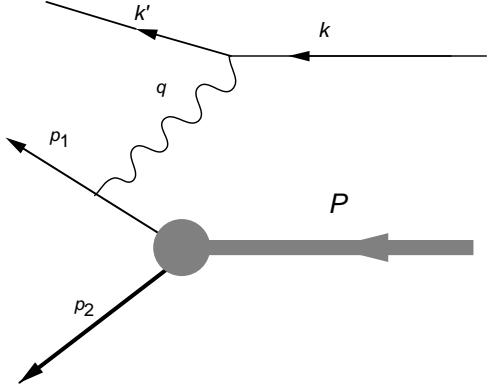}}
\caption{The leading contribution to DIS from a hadron.}\label{nine}
\end{figure}

\section{The nucleon structure functions}

In this section we calculate the cross section and structure functions for 
the deep inelastic scattering (DIS) of unpolarized electrons from the
composite ``nucleon'' $N$,  
\begin{equation}
 e+N \rightarrow e^{ \prime} + X\, ,
\label{4eq1}
\end{equation}
where $X$ is the undetected final hadronic state.  This scattering process
is illustrated in Fig.~\ref{nine}.  The nucleon is composite, as
described in Sec.~II.  For completeness, the well known kinematics and cross
section formula\cite{West,Roberts} are first reviewed in the next two
subsections.  The remaining subsections give results for the 1+1 D toy models
presented in this paper.

\subsection{The Kinematics of DIS}

The Bjorken variable, $x$, was defined in Eq.~(\ref{xdef}) and the invariant
mass of the final hadronic state, $W$, in Eq.~(\ref{4eq3}).  In terms of
these quantities, the energy transferred by the photon to the hadron in the
lab system is
\begin{equation}
q^0= \nu= \frac{Q^2}{2Mx}
\label{4eq2}
\end{equation}
and the magnitude of the spatial component of $q$, which will be chosen to be
in the $\hat z$ direction, is:
\begin{equation}
| {\bf q} |=q_z = \sqrt{Q^2+ \frac{Q^4}{4M^2 x^2}}\, .
\label{4eq1002}
\end{equation}

In the CM of the outgoing hadronic pair, the invariant mass of the final
state is
\begin{equation}
W=\sqrt{m_1^2+ {\bf p}^{\prime 2}_1}+ 
\sqrt{m_2^2+{\bf p}^{\prime 2}_1}\, ,
\label{4eq4}
\end{equation}
where the prime denotes that the variable is evaluated in the CM system.
The magnitude of the three-momentum of the ``quark'' (and ``diquark'') in the
CM frame, and in the Bjorken limit, is
\begin{eqnarray}
{\bf p}^{\prime 2}_1= \frac{\Delta (W^2, m_1^2, m_2^2)}{4W^2} \rightarrow
{Q^2(1-x)\over4x}\left[1 + {x(M^2-2m_1^2-2m_2^2)\over Q^2(1-x)}\right]\, ,
\label{4eq1004}
\end{eqnarray}
where the non-leading term will be needed below.
The components of $q$ in the CM frame can be found by a Lorentz 
boost from the lab:
\begin{equation}
\pmatrix{
q_0^{\prime} \cr 0\cr 0\cr
q_z^{\prime} \cr}= {1\over W}
\pmatrix{
\sqrt{W^2+{q}_z^2} &\; 0\; & \; 0\; & -q_z \cr 0 & 1 & 0 & 0 \cr
0 & 0 & 1 & 0 \cr
-q_z & 0 & 0 & \sqrt{W^2+{q}_z^2} \cr}
\pmatrix{q_0 \cr 0 \cr 0 \cr q_z \cr} \, .
\label{4eq5}
\end{equation}
Hence the CM components are:
\begin{eqnarray}
q_0^{\prime}=&&
\frac{Q^2}{2Mx}\left[M^2+Q^2\left(\frac{1}{x}-1\right)\right]^{-1/2}\left[
\left(M^2+
\frac{Q^2}{x} +\frac{Q^4}{ 4M^2x^2}\right)^{1/2}- 2Mx-\frac{Q^2}{2Mx}\right]
\nonumber\\  
q_z^{\prime}=&&
Q\left[M^2+Q^2\left(\frac{1}{x}-1\right)\right]^{-1/2} 
\sqrt{1+\frac{Q^2}{4M^2x^2}} \nonumber\\
&&\qquad\times\left[\left(M^2+ \frac{Q^2}{x} +
\frac{Q^4}{4M^2x^2}\right)^{1/2}- \frac{Q^2}{2Mx}\right]\, .
\label{4eq6}
\end{eqnarray}
In the deep inelastic limit these become
\begin{eqnarray}
q_0^{\prime}=&& {Q(1-2x)\over 2\sqrt{x(1-x)}} \left[ 1- 
{M^2x\over 2Q^2(1-x)} \right]
\nonumber\\  
q_z^{\prime}=&&{Q\over 2 \sqrt{x(1-x)}} \left[ 1- {M^2x(1-2x)^2\over
2Q^2(1-x)}\right]\, ,
\label{4eq6a}
\end{eqnarray}
where the non-leading terms will be needed later.
Using the same boost, the energy of the initial hadron in the
CM of the outgoing hadronic system is
\begin{eqnarray}
p'_0=&& \frac{M}{W} \sqrt{W^2+{q}^2_z} =
\left[1+{Q^2\over M^2}\left(\frac{1}{x}-1\right)\right]^{-1/2}
\left(M^2+\frac{Q^2}{x} +\frac{Q^4}{ 4M^2x^2}\right)^{1/2}\nonumber\\
\to&& {Q\over 2 \sqrt{x(1-x)}}\left[1-{M^2x[1-4(1-x)]\over
2Q^2(1-x)}\right]=q_z'
\left[1+{M^2\over 2q'^2_z}\right]\, , 
\label{4eq1006}
\end{eqnarray}
as expected.  We now turn to the calculation of the cross section.

\subsection{Cross section and structure functions}

The scattering amplitude for the DIS process is
\begin{equation}
{\cal M}_{fi}= {e^2\over Q^2}\, j_{e\,\mu}(k',k){\cal J}^\mu_a   \, ,
\end{equation}
where ${\cal J}^\mu_a$ is the hadronic current (kept general for now) 
and $j_{e\,\mu}(k',k)$ is the current of the electron
\begin{equation}
j_{e\,\mu}(k',k) = \overline{u}(k's') \gamma_\mu u(k,s) \, .
\end{equation}
The spins of the electron are denoted by $s$ and $s'$, and the momenta are
labeled as in Fig.~\ref{nine}.  In this notation the unpolarized differential
cross section in the lab is
\begin{eqnarray}
\Delta \sigma =&& \int_\Delta\frac{d^3{k}^{\prime}\,d^3 p_1\,d^3 p_2\,} 
{ (2\pi)^9 32 M E E' E_1 E_2} (2\pi)^4\delta(k'+p_1+p_2-k-P)
 |{\cal M}_{fi}|^2 \nonumber\\
=&& {e^4\over Q^4}\,\int_{\Delta} \frac{d^3{k}^{\prime}} { (2 \pi)^3 8 E E'M}
\;\ell^{\mu\nu}\,W_{\mu\nu} \, ,
\label{4eq17}
\end{eqnarray}
where $|{\cal M}_{fi}|^2$ contains a sum over the spins of the final
particles and an average over the spins of the initial particles, $E_1$
and $E_2$ are the energies of the outgoing quark and diquark, $E$ and 
$E^{\prime}$ are the initial and final electron energies, and
$\ell^{\mu\nu}$ and $W^{\mu\nu}$ are the lepton and hadron current tensors,
defined below.  The hadronic tensor contains all of the the physical
information about the hadron-photon interaction and includes the integration
over the  outgoing hadrons and the normalization factors associated with the
hadronic wave functions.  

If the electron mass can be neglected the outgoing
electron differential can be written $d^3k'=E'^2 dE'd\Omega'$, which gives the
following result for the inelastic cross section:
\begin{equation}
\frac{d^2 \sigma}{d \Omega'\, dE^{\prime}}= \frac{\alpha^2}{Q^4}
\left(\frac{E^{\prime}}{E}\right){1\over 4\pi M}\,\ell^{\mu \nu} W_{\mu \nu} 
\, .  \label{4eq10}
\end{equation}
If the electron mass is neglected the lepton current tensor is 
\begin{equation}
\ell_{\mu \nu}= {1\over2} \sum_{\rm spins} j^\dagger_{e\nu} j_{e\mu}
=\frac{1}{2} Tr\left(\gamma_{\nu} \not k^{\prime}\gamma_{\mu}\not k \right)
=2\left(k^{\prime}_{\mu}k_{\nu}+ k_{\mu}k^{\prime}_{\nu} +
\frac{1}{2}\,q^2 g_{\mu \nu}\right) \, .
\label{4eq11}
\end{equation}
Gauge invariance of the hadronic current implies  
\begin{equation}
q^{\mu}W_{\mu \nu}=W_{\mu \nu}q^{\nu}=0\, .
\label{4eq9}
\end{equation}
This in turn implies that the most general form of the hadronic
tensor for unpolarized scattering is
\begin{equation}
{W_{\mu \nu}\over 4\pi M}=-\left(g_{\mu \nu} -
\frac{q_{\mu}q_{\nu}}{q^2}\right) W_1 +
\left(p_{\mu}-q_{\mu} \frac{p \cdot q}{q^2}\right)
\left(p_{\nu}-q_{\nu} \frac{p \cdot q}{q^2}\right)  
\frac{W_2}{M^2}\, ,
\label{4eq9a}
\end{equation}
which {\it defines\/} the structure functions $W_1$ and $W_2$.    

Substituting Eq.~(\ref{4eq9a}) into Eq.~(\ref{4eq10}) gives
\begin{eqnarray}
\frac{d^2 \sigma}{d\Omega'\, dE^{\prime}}=&& \left(\frac{2\alpha E'}{Q^2}
\right)^2\,\left[W_2 \,\cos^2 \left(\frac{\theta}{2}\right) +
2W_1\,\sin^2 \left(\frac{\theta}{2}\right)\right] \nonumber\\
=&& \sigma_M\,\left[W_2  +
2W_1\,\tan^2 \left(\frac{\theta}{2}\right)\right] \, ,
\label{4eq12}
\end{eqnarray}
where $\theta$ is the scattering angle of the electron in the lab system and
$\sigma_M$ is the Mott cross section.

We now calculate the hadronic tensor.  If the spin of the target is $S$,
in 1+3 dimensions this tensor is  
\begin{eqnarray}
W_{\mu \nu}(q,p)=&& \frac{1}{(2S+1)} \sum_{S,s_1,s_2}
\int{d^3p_1'd^3p_2'\over  (2\pi)^6
4E'_1E'_2}(2\pi)^4\delta^4(p_1+p_2 -q -p) {\cal J}^\dagger_{a\nu}
{\cal J}_{a\mu}\nonumber\\
=&&\frac{1}{(2S+1)}\sum_{S,s_1,s_2}\; {|{\bf p}'_1|\over  4\pi
(E'_1+E'_2)}\,{1\over4\pi}\, \int d\Omega'_1 \; 
{\cal J}^\dagger_{a\nu} {\cal J}_{a\mu}\, .
\label{4eq7}
\end{eqnarray}
where $S$, $s_1$, $s_2$ are the spins, and the second line gives the result
in the CM system.  The volume integrals $d^3p_1/E_1$ and
$d^3p_2/E_2$ are covariant, insuring that the tensor is also covariant. 
However, when we treat the hadronic degrees of freedom in 1+1
dimensions, consistency requires that the hadronic tensor also be evaluated
in 1+1 dimensions.  In this case Eq.~(\ref{4eq7}) must be replaced by its
1+1 dimensional equivalent
\begin{eqnarray}
W'_{\mu \nu}(q,p)=&& \frac{1}{(2S+1)} \sum_{S,s_1,s_2}
\int{dp_1'dp_2'\over  (2\pi)^2
4E'_1E'_2}(2\pi)^2\delta^2(p_1+p_2 -q -p) {\cal J}^\dagger_{a\nu}
{\cal J}_{a\mu} \nonumber\\
=&&\frac{1}{(2S+1)} \sum_{S,s_1,s_2}\frac{1}{4 |{\bf p}'_1|(E'_1+E'_2)}
\;\Bigl\{{\cal J}^{\dagger}_{a\nu}(+) {\cal J}_{a\mu}(+) + {\cal
J}^{\dagger}_{a\nu}(-) {\cal J}_{a\mu}(-) \Bigr\} \, ,\nonumber\\
&&
\label{4eq7a}
\end{eqnarray}
where the second line gives the result
in the CM system with ${\cal J}_{a\mu}(\pm)$ corresponding to the cases when
${\bf p}'_1$ is parallel to ${\bf q}$ (+) or antiparallel ($-$).  Except for
these two possibilities the delta functions completely fix the kinematics.

To calculate the structure functions for the 1+1 dimensional models, we use
Eqs.~(\ref{3eq23}), (\ref{5eq1b}), and (\ref{4eq7a}).  The definitions
(\ref{4eq7a}) and (\ref{4eq9a}) are covariant, so the structure functions can
be evaluated in any frame, and it is convenient to evaluate them in the CM
frame of the outgoing hadrons. Furthermore, since $q'$ and $p'$ have
components only in the $\hat t$ and
$\hat z$ directions, gauge invariance insures that 
\begin{eqnarray}
W^{0z}=W^{z0}={q'_0\over q'_z}W^{00}={q'_z\over q'_0}W^{zz}
\end{eqnarray}
and that $W^{yy}=W^{xx}$.  Hence $W_1$ and $W_2$ can be extracted from
\begin{eqnarray}
W^{xx}=&& 4\pi M\,W_1 \nonumber\\
W^{00}-W^{zz}=&& 4\pi M\,\left[ \left(1+{Q^2 \over
4 M^2 x^2}\right) W_2 -W_1\right]\, . \label{eqcg}
\end{eqnarray}
In the next two subsections we calculate the structure functions for the
scalar and spin 1/2 models in 1+1 dimensions.

\subsection{Structure functions for the scalar model}

Using Eqs.~(\ref{3eq23}) and (\ref{4eq7a}), the structure functions for the
scalar theory in the deep inelastic limit are
\begin{eqnarray}
W^{xx}=&& 0 \nonumber\\
W^{00}-W^{zz}=&& \frac{2x\;{\cal N}^2}{Q^4(1-x)}\;\left\{\frac{(q\cdot
p_1)^2}{(2q\cdot p_1 - q^2)^2}\,\Biggl|_{+} + \frac{(q\cdot p_1)^2}{(2q\cdot
p_1 - q^2)^2}\,\Biggl|_{-} \right\}
\, , 
\end{eqnarray}
where
\begin{eqnarray}
(q\cdot p_1)^2|_+=&&(q_0' E_1'-q'_zp'_1)^2=
{\textstyle\frac{1}{4}}{Q^4}\nonumber\\ 
(q\cdot p_1)^2|_-=&&(q_0' E_1'+q'_zp'_1)^2={\textstyle\frac{1}{4}}
{Q^4}\;\left(\frac{1-x}{x}\right)^2\nonumber\\
(2q\cdot p_1 - q^2)|_+=&& b-a \nonumber\\ (2q\cdot p_1 - q^2)|_-=&& b+a  \, , 
\end{eqnarray}
and
\begin{eqnarray}
a=&&2{p}_1^{\prime} q'_z \rightarrow  {Q^2 \over 2 {x}} 
\left[1-{x\left(m_1^2+m_2^2-2M^2x(1-x)\right)\over Q^2(1-x)}
\right]\nonumber\\
b=&&Q^2 + 2E'_1q^{\prime}_ 0 \rightarrow Q^2 + {Q^2 (1-2x)\over 2 {x}}
\left[1-{x(m_2^2-m_1^2)\over Q^2(1-x)}\right]
\nonumber\\
\rightarrow &&{Q^2 \over 2 {x}} \left[1- {x(1-2x)(m_2^2-m_1^2)\over
Q^2(1-x)} \right]\, .
\label{4eq18}
\end{eqnarray}
Hence $b+a\to Q^2/x$, and the $(-)$ term vanishes in the deep inelastic
limit.  The  (+) term is finite however, giving 
\begin{eqnarray}
W_1 \rightarrow&& 0 \nonumber\\
\nu W_2\rightarrow&& {x\over2\pi}(W^{00}-W^{zz})
\rightarrow \frac{{x^2\;\cal N}^2}{4\pi(1-x)}\; \left(\frac{1}{b-a}\right)^2
\nonumber\\
\rightarrow&& \frac{{\cal N}^2}{4\pi}\;\frac{x^2\,(1-x)}{[m_1^2(1-x)+
m_2^2x-M^2x(1-x)]^2} 
\, . \label{W2}
\end{eqnarray}
$W_1$ vanishes in the deep inelastic limit, as expected for scalar
particles. But $\nu W_2$ {\it scales\/} in the deep inelastic region; 
its dependence on $Q^2$ and $\nu$ is replaced by dependence
on the Bjorken scaling variable $x$ alone.  In the simple quark parton
model, we expect
\begin{eqnarray}
\nu W_2(Q^2,\nu) \rightarrow F_2(x) = xf(x)\, ,
\label{4eq13}
\end{eqnarray}
where $f(x)$ is the probability of finding a quark with momentum fraction $x$
\begin{eqnarray}
\int_0^1dx\,f(x)=1\, .
\label{4eq13a}
\end{eqnarray}
Our simple model therefore predicts
\begin{eqnarray}
f(x)=\frac{{\cal N}^2}{4\pi}\;\frac{x\,(1-x)}{[m_1^2(1-x)+
m_2^2x-M^2x(1-x)]^2} \, ,
\end{eqnarray}
which satisfies the normalization condition (\ref{4eq13a}), as can
be seen by comparing with Eq.~(\ref{1eqlast}).

\subsection{Structure functions for the spin 1/2 model} 

In the previous subsection we obtained the familiar result that the
Callan-Gross relation\cite{CG} is not satisfied by scalar quarks.  This
relation predicts that   
\begin{eqnarray} 
 MW_1(Q^2,\nu) \rightarrow F_1(x)={\textstyle\frac{1}{2}}f(x) \, ,
\label{4eq14}
\end{eqnarray}
where  $f(x)$ the same function which specified $F_2$ in Eq.~(\ref{4eq13}).
For scalar quarks the structure function $W_1$ falls off with $Q^2$. To obtain
the relation (\ref{4eq14}), the quarks must to be treated as spin 1/2
fermions.  [The diquarks, even when their orbital angular momentum is zero,
will have two spin states (0 and 1), and our choice of spin zero corresponds
to the simplest possible wave function.]  When carrying out the calculation
in 1+1 D, we must assume that only the momenta are restricted to 1+1 D, but
that the spin lives in the full 1+3 dimensional space.  This assumption is
fully consistent with the dynamical calculations of Secs.~II and III.    

From the spin zero calculation we know that it is sufficient to evaluate the
(+) matrix element only.  In this case the incoming nucleon is traveling in
the $-\hat z$ direction, and the final quark in the $+\hat z$ direction.  We 
will calculate this matrix element in the CM system of the outgoing pair
using helicity states, which for this case are
\begin{eqnarray}
{U}_N(p, -\lambda_N)=&& \sqrt{E_p+M} \pmatrix{1 \cr -2 \lambda_N {\tilde
p} \cr} \chi_{_{\lambda_N}} \nonumber\\
u(p_1, \lambda _1)=&&\sqrt{E_{p_1}+m_1} \pmatrix{1 \cr
2 \lambda_1 {\tilde p}_1 \cr} \chi_{_{\lambda_1}}
\label{5eq3}
\end{eqnarray}
where 
\begin{eqnarray}
{\tilde p}=&&\frac{p}{E_p+M} \simeq 1-\frac{M}{p}\nonumber\\
{\tilde p}_1=&&\frac{p_1}{E_{p_1}+m_1}\simeq 1-\frac{m_1}{p_1} \nonumber\\
\chi_{+}=&&\pmatrix{1 \cr 0 \cr}\qquad
\chi_{-}=\pmatrix{0 \cr 1 \cr} \, .
\end{eqnarray}
With the help of these quantities, the $x$-component of the
current (\ref{5eq1b}) becomes:
\begin{eqnarray}
{\cal J}_{a}^x=&&\sqrt{(E_{p_1}+m_1)(E_p+M)}\,\frac{{\cal N}}{b-a}
\; 2 \lambda_N\,[q^0({\tilde
p}_1-{\tilde p})+q^3({\tilde p}_1{\tilde p}-1)]\, \delta_{\lambda_N,
\lambda_1}\nonumber\\
\rightarrow&& -\frac{{\cal N} Q}{\sqrt{x}\,[b-a]}\;2\lambda_N \,[Mx+m_1]\, .
\label{5eq4}
\end{eqnarray}
This gives
\begin{eqnarray}
W_1= {{\cal N}^2\over 8 \pi M }\; \frac{(1-x) (Mx+m_1)^2}{[m_1^2(1-x)+
m_2^2x-M^2x(1-x)]^2} = {\frac{1}{2M}}f(x)
\label{5eq4a}
\end{eqnarray}
In a similar manner, it may be shown that $W^{00}-W^{zz}\to0$ in
the deep inelastic limit, which is the Callan Gross relation [recall
Eq.~(\ref{eqcg})].

Note that the normalization condition for the spin 1/2 model given in
Eq.~(\ref{1eq27}) confirms that this model is also correctly normalized.  
However, the $f(x)$ obtained from the spin 1/2 model does not go to zero as
$x\to0$.

\subsection{Numerical results}

\begin{figure}
\centerline{\epsfxsize=4.5in\epsffile{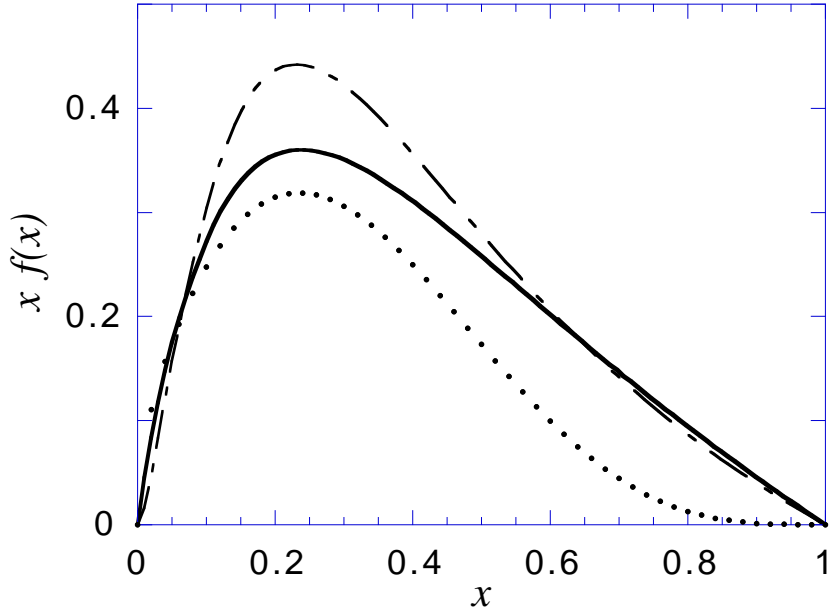}}
\caption{Quark momentum distributions:  $xu(x)$ with $u(x)$
normalized to unity (dotted line), $xf(x)$ for the scalar model
with $m_1^2/M^2=0.1$ and  $m_2^2/M^2=2$ (dashed line), and $xf(x)$ for the 
spin 1/2 model with $m_1^2/M^2=0.18$ and  $m_2^2/M^2=2$ (solid
line).}\label{ten}
\end{figure}

In Fig.~\ref{ten} the structure functions obtained from the scalar and
spin 1/2 models are compared with the empirical momentum distribution $xu(x)$ 
for $u$ quarks in the proton.  The empirical distribution was taken
from Ref.~\cite{dist} and has been renormalized to unity for
comparison.

Note that the parameters in the momentum distributions predicted by the models
can be chosen to give a qualitatively reasonable description of empirical
results, but the models fail at both large and small $x$.  The large
$x$ distributions fall like $1-x$, while the empirical distribution falls
more rapidly (the fit gives $(1-x)^{3.509}$ compared to $(1-x)^3$ for the
naive quark model).  Similarly, the empirical distribution goes
like $x^{0.4810}$ at small $x$, while the spin 1/2 model goes like
$x$ and the scalar model goes like $x^2$.  Assuming the $u$ and $d$ quark
distributions are identical, the resulting momentum carried by valence quarks
is $3\times0.150=0.450$ for the empirical distribution,  $3\times0.210=0.630$
for the spin 1/2 model, and  $3\times0.232=0.696$ for the scalar
model.  In every case some momentum must be carried by glue or
sea quarks, and the naive models presented here do not include
these contributions.  This will be investigated in a future work.            

\section{CONCLUSIONS}

In this paper we have shown how to extract the  gauge invariant part of the
pole term so that it gives the {\it leading contribution to deep 
inelastic scattering\/}.  The gauge non-invariant part depends only on
$q^\mu$ and is cancelled exactly by the rescattering term.  While we have
not shown it explicitly, the argument suggests that this procedure is quite
general and works as follows.  First add a (unique) term proportional to
$q^\mu$ to the pole term which makes it gauge invariant.  Then subtract
the same term from all of the remaining terms (rescattering and
interaction currents).  It follows that the remaining terms are also
gauge invariant, and we {\it conjecture\/} that the sum of all of these
remaining terms will vanish in the deep inelastic limit.  Hence the modified
pole term will give the {\it exact\/} result for DIS.

More explicitly, imagine that the full current is the sum of
a pole term $J_{\rm pole}^\mu$ and a remainder $J_{\rm rem}^\mu$, and that it
is gauge invariant.  The first step is to write it as
\begin{eqnarray}
J^\mu=&&J_{\rm pole}^\mu+J_{\rm rem}^\mu\nonumber\\
=&&\left(J_{\rm pole}^\mu -\alpha\, q^\mu\right) +\left(J_{\rm rem}^\mu + 
\alpha\,q^\mu\right) \, ,
\end{eqnarray}
where 
$$\alpha=\frac{J_{\rm pole}\cdot q}{q^2}\, .$$
Note that $\alpha$ is uniquely determined (unless $q^2=0$, which cannot
happen for DIS), and that the modified pole term
\begin{eqnarray}
J_{\rm mod pole}^\mu=J_{\rm pole}^\mu- \frac{J_{\rm pole}\cdot
q}{q^2}\;q^\mu \label{last}
\end{eqnarray}
is gauge invariant by construction.  We conjecture that the modified
remainder terms (which are also gauge invariant) will vanish in the DIS
limit, leaving (\ref{last}) to give the exact result for DIS.  Note
that our prescription (\ref{last}) is identical to the Landau
prescription, and defined in Ref.~\cite{Kelly}.

The second result of this paper is the construction of toy models for
the discription of DIS.  These models give a qualitatively reasonable
description of the phenomenology, as shown in Fig.~\ref{ten}, but
simplicity is their main virtue.  We plan to use them to study many
issues in the theory of DIS.

\acknowledgments

We would like thank Richard Lebed for calling our attention to
Ref~\cite{Einhorn}, and Christian Wahlquist for supplying the empirical quark
distribution functions used in Fig.~\ref{ten}.  The support of the DOE
through grant  No.~DE-FG02-97ER41032 is gratefully acknowledged.


\begin{references}

\bibitem{West} G.~W.~West, Phys.~Rep.~{\bf 18}, 263 (1975).

\bibitem{Roberts} R.~G.~Roberts, {\it The Structure of the Proton\/} (1990).

\bibitem{covardeut}  F.~Gross and S.~Liuti, Phys.~Rev.~C {\bf 45}, 1374
(1992); A.~Yu.~Umnikov and F.~C.~Khanna, Phys.~Rev.~C {\bf 49}, 2311
(1994); W.~Melnitchouk, A.~W.~Schreiber, and A.~W.~Thomas, Phys. Lett. {\bf
B335}, 11 (1994); W.~Melnitchouk, G.~Piller, and A.~W.~Thomas, Phys. Lett.
{\bf B346}, 165 (1995). 

\bibitem{df83} T.~de Forest, Jr.~Nucl.~Phys.~{\bf A392}, 232 (1983);
Ann.~Phys.~(N.Y.) {\bf 45}, 365 (1967).

\bibitem{Kelly} J.~Kelly, Phys.~Rev.~C {\bf 56}, 2672 (1997).

\bibitem{Einhorn} M.~Einhorn, Phys.~Rev.~D {\bf 14}, 3451 (1976).

\bibitem{GR} F.~Gross and D.~O.~Riska, Phys.~Rev.~C {\bf 36}, 1928 (1987).

\bibitem{CG} C.~Callan and D.~ Gross, Phys.~Rev.~Lett.~{\bf 22}, 156 (1969).

\bibitem{dist} J.~C.~Collins and Wu-Ki Tung, Nucl.~Phys.~{\bf B278}, 934
(1986).  These fits were provided to us by C.~Wahlquist.

\end{references}
\end{document}